%% file: XMM_Centaurus_paper.tex
\documentclass[useAMS,usenatbib]{mn2e}
\usepackage{amssymb,amsmath,epsfig,times, natbib, color}

\title[Merging in Centaurus]{An \emph{XMM-Newton} view of the merging activity in the Centaurus cluster}
\author[S. A. Walker et al.]{S. A. Walker,$^1$\thanks{Email: 
    swalker@ast.cam.ac.uk} A. C. Fabian,$^1$ J. S. Sanders$^2$ \\
  $^1$Institute of Astronomy, Madingley Road, Cambridge CB3 0HA \\
  $^2$Max-Planck-Institute fur extraterrestrische Physik, 85748 Garching, Germany \\
    \\
   \\
   \\
}
\date{}

\begin{document}

\maketitle

\begin{abstract}
 We report the results of \emph{XMM-Newton} observations of the regions around the core of the Centaurus cluster where evidence for merging activity between the subgroup Cen 45 and the main Centaurus cluster has previously been observed using \emph{ASCA} and \emph{ROSAT} data. We confirm the \emph{ASCA} findings of a temperature excess surrounding Cen 45. We find that this temperature excess can be explained using simple shock heating given the large line of sight velocity difference between Cen 45 and the surrounding main Centaurus cluster. We find that there is a statistically significant excess in metallicity around Cen 45, showing that Cen 45 has managed to retain its gas as it has interacted with the main Centaurus cluster. There is a pressure excess to the east in the direction of the merger, and there is also an entropy excess around the central galaxy of Cen 45. The metallicity between 50-100 kpc to the north of NGC 4696 is higher than to the south, which may be the result of the asymmetric distribution of metals due to previous sloshing of the core, or which may be associated with the filamentary structure we detect between NGC 4696 and NGC 4696B.
\end{abstract}


\section{Introduction}

\begin{figure*}
  \begin{center}
    \leavevmode
      \epsfig{figure=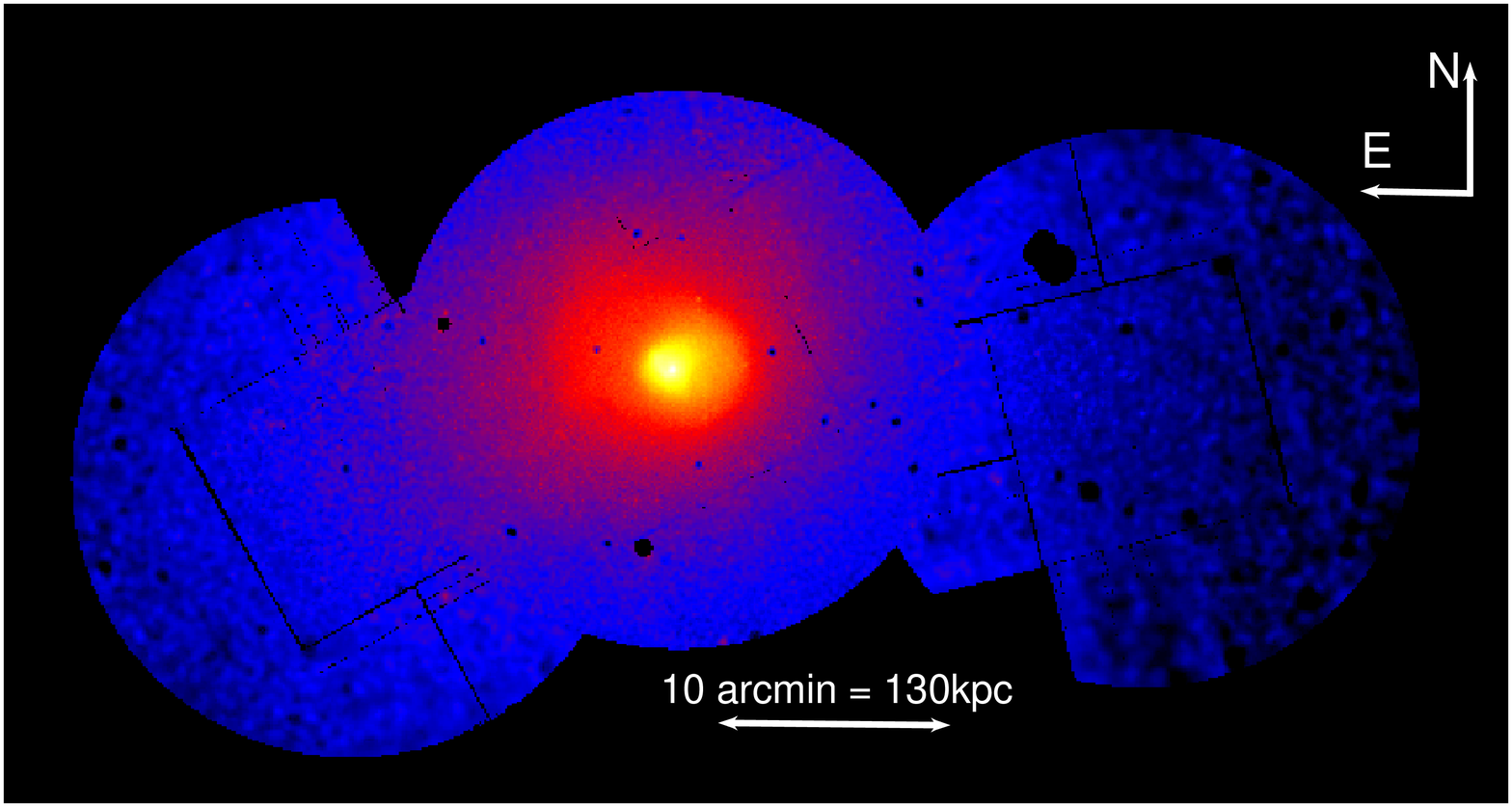,
        width=0.49\linewidth}
         \epsfig{figure=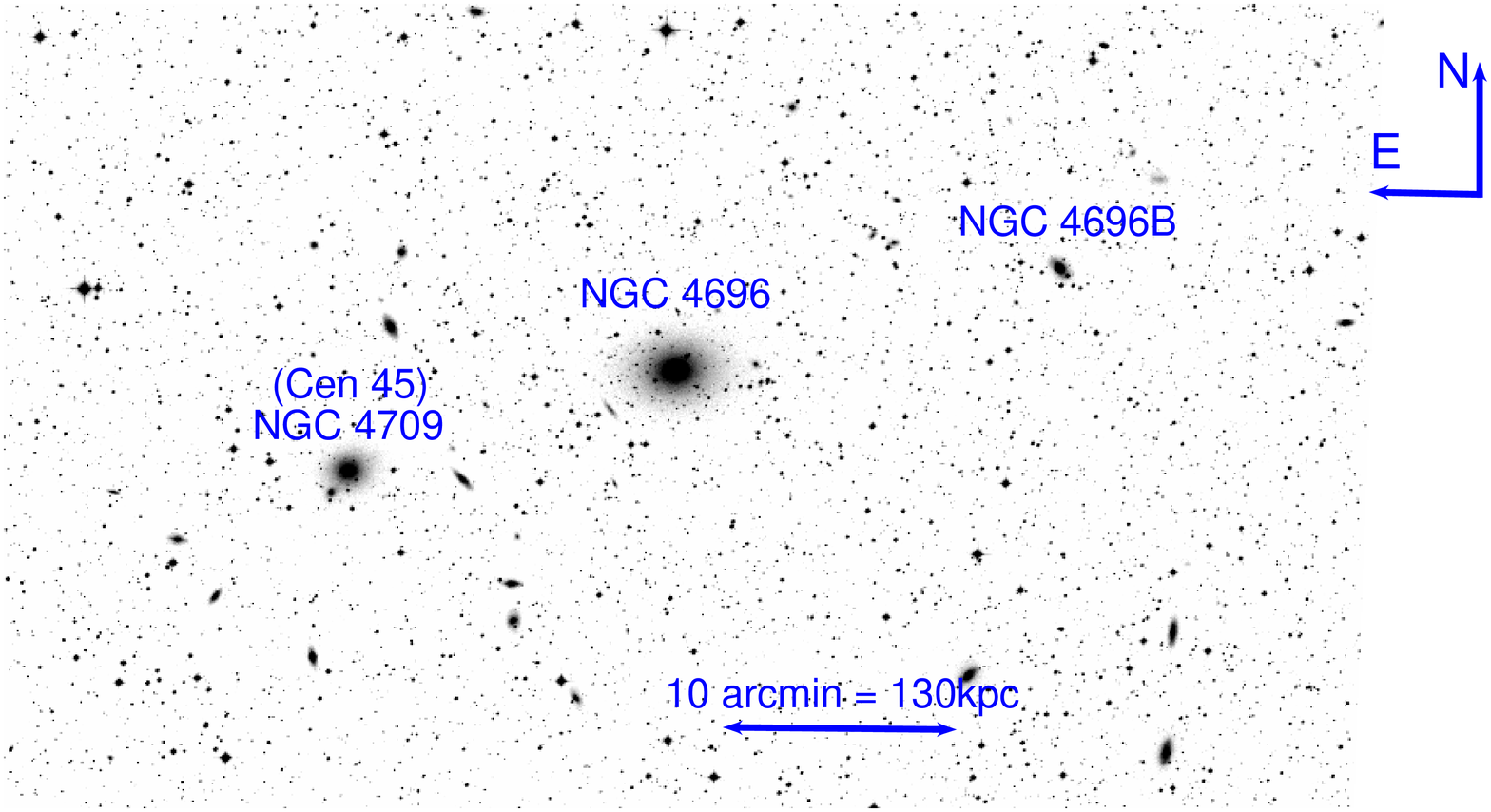,
        width=0.49\linewidth}     
        
      \caption{\emph{Left}:Exposure corrected, background subtracted, point source subtracted and adaptively smoothed mosaic X-ray image in the 0.7-7.0 keV band produced as described in section \ref{imageproduction}. \emph{Right}: Digitised Sky Survey (DSS) image with main galaxies labelled, and with the coordinates set to match the X-ray image in the left panel. }
      \label{images}
  \end{center}
\end{figure*}

Optical observations of the Centaurus cluster \citep{Lucey1986} have shown there to be two distinct clusterings of galaxies. The main one associated with the main Centaurus cluster, centred on the brightest cluster galaxy (BCG) NGC 4696 (Fig. \ref{images} right panel) has a line of sight velocity of 3000 km s$^{-1}$ and is referred to as Cen 30. A smaller subgroup containing the second brightest galaxy in Centaurus (NGC 4709) is located 15 arcmins (190kpc) to the east and contains a factor of 2.5 fewer galaxies, and has a significantly lower velocity dispersion (280 km s$^{-1}$) than the main Centaurus cluster (586 km s$^{-1}$). This smaller subgroup also has a higher line of sight velocity of 4500 km s$^{-1}$, and hence it is referred to as Cen 45. Despite this large line of sight velocity difference (1500 km s$^{-1}$), \citet{Lucey1986} found that luminosity functions, colour-magnitude relations and galactic radius distributions indicate that both Cen 30 and Cen 45 are at a common distance. 

\citet{Churazov1999} used \emph{ROSAT} and \emph{ASCA} data and found there to be a temperature enhancement associated with Cen 45, around 20 arcmins to the south east of the core of the main Centaurus cluster. This temperature enhancement was also found in the analysis of the same \emph{ASCA} data in \citet{Furusho2001} and \citet{Dupke2001}. If the ICM emission is purely the result of the superposition of the Centaurus cluster and Cen 45, we would instead expect a temperature decrement as the less massive Cen 45 should contain gas of a lower temperature. The X-ray data therefore add further strong evidence that Cen 30 and Cen 45 lie at the same distance and are interacting. \citet{Churazov1999} concluded that this temperature excess indicates that the ICM has been heated as Cen 45 merges with the main Centaurus cluster. However, due to the large PSF of \emph{ASCA}, the temperature structure could not be well resolved. 

Here we report \emph{XMM-Newton} observations of the Cen 45 system and the regions around the core of the Centaurus cluster, which have allowed the temperature structure of the system to be spatially resolved, whilst also allowing the metallicity structure to be explored. The Centaurus galaxy cluster is the second closest cluster (z=0.0104 corresponds to a distance of 44 Mpc and an angular scale of 13 kpc/arcmin), and has been studied extensively with all major X-ray observatories. Centaurus' large angular extent makes it an excellent target for highly spatially resolved studies of the intracluster medium and any ongoing merging activity.  

Deep \emph{Chandra} observations of the core (the central 40 kpc) of the Centaurus cluster \citep{Fabian2005, Sanders2006} have found a significantly asymmetric temperature and metallicity structure around NGC 4696. The cold, high metallicity gas in the core is displaced to the west of the BCG.  There are two sharp surface brightness discontinuities around the core, one 17.5 kpc to the east centred on NGC 4696 and one 32.1 kpc to the west and centred 10.9kpc to the west of the core, suggesting an east-west motion of the gas in the cluster potential. These sharp surface brightness discontinuities coincide with abrupt temperature changes, suggesting that they are cold fronts produced by the sloshing motion of the cool core in the cluster potential relative to the surrounding hotter ICM. It is possible that this sloshing motion of the core has been caused by the passage of Cen 45 through the main cluster. 

Observations with \emph{Suzaku} out to 140kpc to the north and south of the core \citep{Ota2007} have found no significant bulk motions of the intracluster medium, suggesting that in the central regions the bulk velocity does not excees the thermal velocity of the gas.

\section{Observations and Data Reduction}

A mosaic of 4 \emph{XMM} MOS observations was studied (Fig. \ref{images} left panel), and these observations are summarised in table \ref{obsdetails}. Two pointings exist for the centre, centred on the BCG of Centaurus (NGC 4696). A pointing to the south east is centred on the second brightest cluster galaxy (NGC 4709) which is the dominant galaxy in the Cen 45 subgroup. The final pointing is to the south west. 
The observations were reduced and analysed using the Extended Source Analysis Software (ESAS version 12) as described in \citet{Snowden2008}, following the 'Cookbook for analysis procedures for XMM-Newton EPIC MOS observations of extended objects and the diffuse background'\footnote{ftp://legacy.gsfc.nasa.gov/xmm/software/xmm-esas/xmm-esas.pdf}. We used only the MOS data here, but in section \ref{PN_data} we use the PN data as an independent test of the MOS results, finding completely consistent results. The source to background ratio in the band used for spectral fitting (0.7-10.0keV) decreases from 80 at the centre to 1.5 in the most peripheral bin for the MOS data.

We ran \textsc{emchain} and \textsc{mos-filter} on the MOS data to remove periods of high soft proton flux. CCD 6 of MOS1 was not available (due to a micrometeorite hit) for all of the observations except for the shorter central observation, which was taken before the micrometeorite hit. CCDs in the anomalous state were removed (these were CCDs 4 and 5 for the MOS1 eastern pointing, CCDs 4 and 5 for the MOS1 western pointing, CCD 5 for the MOS2 western pointing, CCDs 4 and 5 for the MOS1 long central pointing and CCD 5 for the MOS2 long central pointing). 

To check for residual soft proton contamination we compared the 10-12 keV count rate per unit area in the actual field of view (where the mirrors focus the soft protons) with the same count rate in the unexposed corners (which are not affected by soft protons). Since the source contribution in this hard band is negligible, this allows the identification and removal of periods of significant residual soft proton contamination \citep{Leccardi2008}. In table \ref{inoverout} we show the "IN over OUT" diagnostic \citep{DeLuca2004} for the observations, which is the ratio of the surface brightness in the 6-12 keV band in an outer region of the field of view (the IN component) to the surface brightness in the same band calculated outside the field of view (the OUT component). All of the values are only slightly above 1.0, indicating negligible contamination.
 
Point sources were identified and excised using the task \textsc{cheese} down to a uniform threshold flux of 1$\times$10$^{-14}$ erg cm$^{-2}$ s$^{-1}$ in the 0.7-7.0 keV band to ensure that the cosmic X-ray background is resolved uniformly over all of the observations. 
 
Spectra for the regions shown in Fig. \ref{TandZmaps} (excluding the point sources identified earlier) were extracted using the task \textsc{mos-spectra}, which also created the ARFS and RMFs. The quiescent particle background (QPB) for each region was obtained using the task \textsc{mos-back}. 
 
When performing the spectral extraction, the data were binned so that each spatial region contained the same number of counts (20,000 in the 0.7-7.0 keV band), allowing robust measurements of both the temperature and metallicity to be obtained. The regions were determined using Voronoi tessellation method of \citet{Diehl2006}, which has the advantage that it produces geometrically unbiased regions which do not lead the eye, and which is commonly used for binning \emph{XMM} data (e.g. \citealt{Simionescu2007}). This is particularly important because the central pointing has over 4 times the exposure of the offset pointings, while some of the MOS chips were in the anomalous mode and had to be ignored, so the raw total counts distribution when both detectors are summed is very uneven compared to actual X-ray emission, and could lead to misleading geometric features if alternative binning techniques were used. 

For the central regions (the central 3 arcmins) of the temperature and metallicity maps shown later in Fig. \ref{TandZmaps}, we show the Chandra results from \citet{Sanders2006}, which have superior spatial resolution than the \emph{XMM-Newton} results. 

As in \citet{Urban2011}, we use only the data from within the central 12 arcmin radius region for each MOS observation (centred on the aim point) to minimise the systematic uncertainties associated with the effective area calibration at large off-axis angles.

\subsection{Image Production}
\label{imageproduction}
Images and exposure maps each detector for each pointing in the 0.7-7.0 keV band were extracted using \textsc{mos-spectra}. Particle background images were then produced using \textsc{mos-back} for each detector. The soft proton background images for each detector for each pointing were then created using the task \textsc{proton}, using the index and normalisation for the soft proton broken powerlaw component fitted in the background model later in section \ref{newbackgroundmodelling} as inputs. To model the soft proton contribution we used a broken powerlaw model with a slope of 0.4 below 5 keV and a slope of 0.8 above 5 keV (as in \citet{Leccardi2008}). The soft proton images and the particle background images were rotated from detector coordinates to sky coordinates using the task \textsc{rot-im-det-sky}. The final background subtracted, point source removed and exposure corrected mosaic shown in Fig. \ref{images} (left panel) was then obtained using the task \textsc{merge\_comp\_xmm}. The image was then smoothed with the task \textsc{adapt\_2000} using smoothingcounts=50 and thresholdmasking=0.02.

\subsection{Hard band image, pressure map}
\label{hardbandimageproduction}

The observed count rate from the ICM can be related to its pressure, $P$,  and temperature, $kT$, as follows (e.g. \citealt{Forman2007}):
\begin{equation}
C \propto \int P^2 \epsilon(T)/T^2 dl
\label{eq:px}
\end{equation}
where $\epsilon(T)$ is the volume emissivity of the ICM convolved with the \emph{XMM-Newton} MOS response, and $l$ is the line of sight path length. In the 3.5-7.5 keV band the quantity $\epsilon(T)/T^2$ is only very weakly dependent on the gas temperature in the temperature range of interest (1-4keV), as shown in Fig. \ref{hardbandindependence}. This means that the image in the 3.5-7.5 keV band is essentially a map of $\int P^2 dl$, the square of the pressure integrated along the line of sight. 

\begin{figure}
  \begin{center}
    \leavevmode
      \epsfig{figure=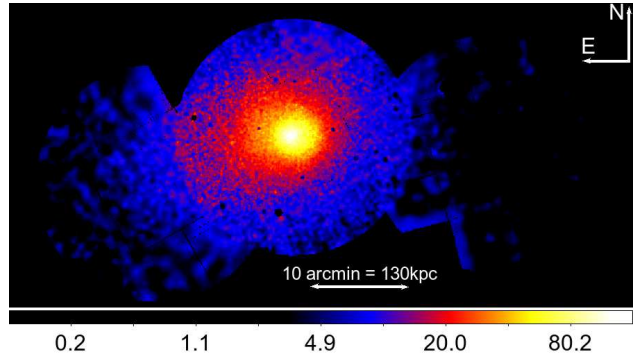,
        width=\linewidth}

      \caption{Exposure corrected, background subtracted, point source subtracted and adaptively smoothed mosaic X-ray image in the hard band (3.5-7.5 keV) which maps the pressure distribution, showing the excess in pressure to the east. }
      \label{pressuremap}
  \end{center}
\end{figure}

The hard band image pressure map is shown in Fig. \ref{pressuremap}, where we see a clear excess to the east in the direction of the Cen 45 merger.  

\begin{table}
  \begin{center}
  \caption{Observational parameters of the pointings}
  \label{obsdetails}
  
    \leavevmode
    \begin{tabular}{llllll} \hline \hline
    Obs. ID & Position & Total exposure  & RA & Dec (J2000)  \\ 
         &  & per detector (ks) &  &  \\ \hline
    0046340101 & Centre & 47 &192.2 &-41.3 \\ 
     0406200101   & Centre & 124 & 192.2&-41.3\\ 
     0504360101 & East & 43 &192.5 &-41.4\\ 
      0504360201  & West & 34 &191.7 &-41.3 \\ \hline

    \end{tabular}
  \end{center}
\end{table}

\begin{table}
  \begin{center}
  \caption{"In over out" ratios for the observations in the 6-12 keV band as described in the main text.}
  \label{inoverout}
  
    \leavevmode
    \begin{tabular}{llll} \hline \hline
    Obs. ID & Position &  "In over out" ratio \\ 
         &  & MOS1 & MOS2\\ \hline
    0046340101 & Centre & 1.15&1.12\\ 
     0406200101   & Centre &1.10 &1.07 \\ 
     0504360101 & East &1.14 &1.11\\ 
      0504360201  & West &1.12  &1.08\\ \hline

    \end{tabular}
  \end{center}
\end{table}

 \begin{table}
  \begin{center}
  \caption{Soft foreground measurements from \emph{ROSAT} RASS data as described in the text. The units of the \textsc{apec} normalisations are $10^{-14}(4\pi)^{-1}D_{A}^{-2}(1 + z)^{-2} \int n_{e} n_{H} dV$, where $D_{A}$ is the angular size distance (cm), and $n_{e}$ and $n_{H}$ are the electron and hydrogen densities (cm$^{-3}$) respectively, and these values are scaled for a circular area of sky of 1 arcmin$^{2}$}
  \label{GALvariations}

    \leavevmode
    \begin{tabular}{ll} \hline \hline

           0.22 keV \textsc{apec} norm (GH) & 0.12 keV \textsc{apec} norm (LHB)    \\ \hline

     1.8$^{+0.3}_{-0.2}$ $\times$ 10$^{-6}$ & 8.0$^{+0.8}_{-0.8}$ $\times$ 10$^{-7}$ \\ \hline

    \end{tabular}
  \end{center}
\end{table}

\begin{figure*}
  \begin{center}
    \leavevmode
    
    \hbox{
      \epsfig{figure=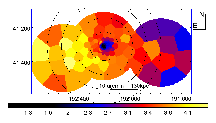,
        width=0.5\linewidth}
         \epsfig{figure=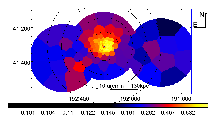,
        width=0.5\linewidth}     
        }
        
       \hbox{
       \epsfig{figure=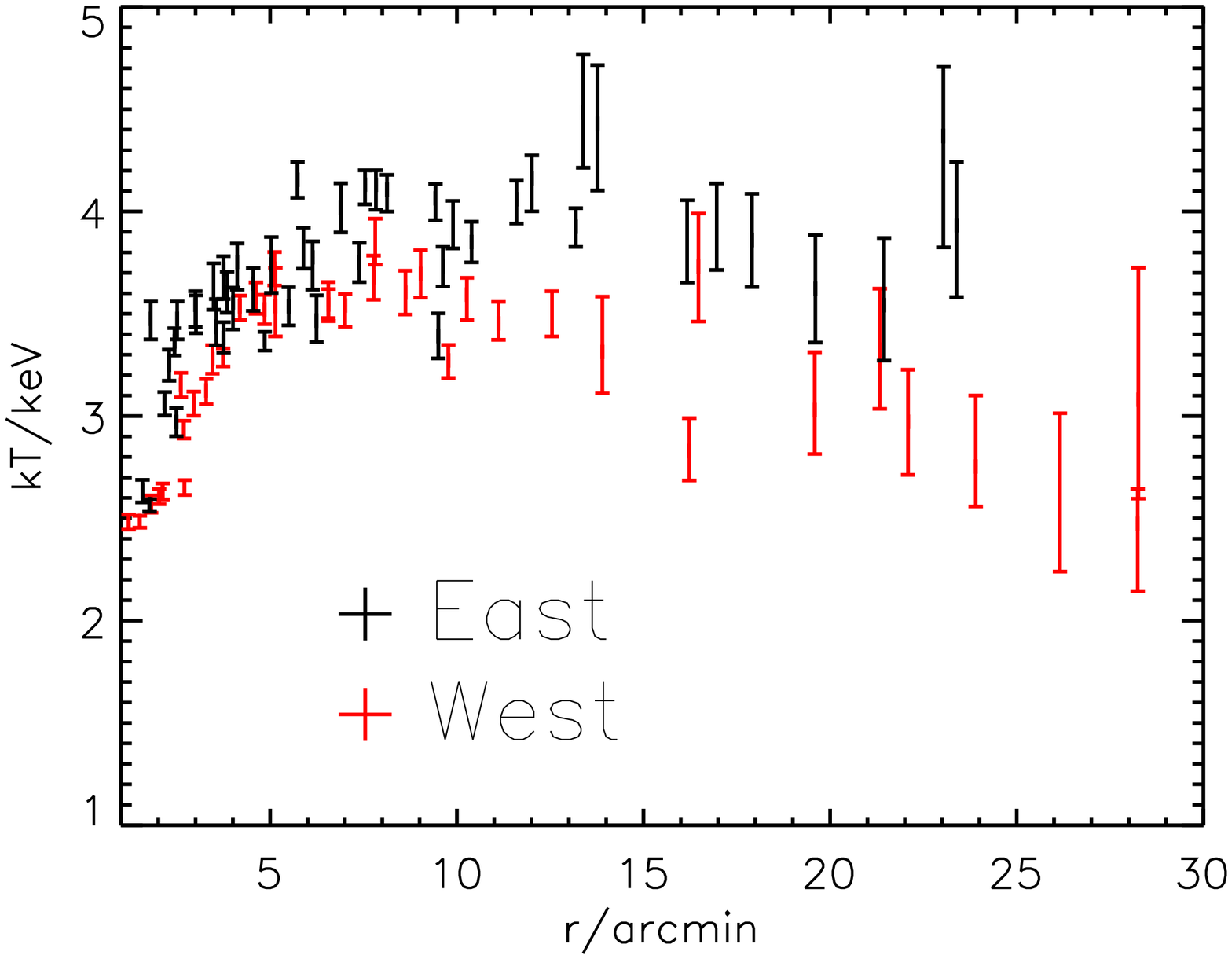,
        width=0.5\linewidth}
         \epsfig{figure=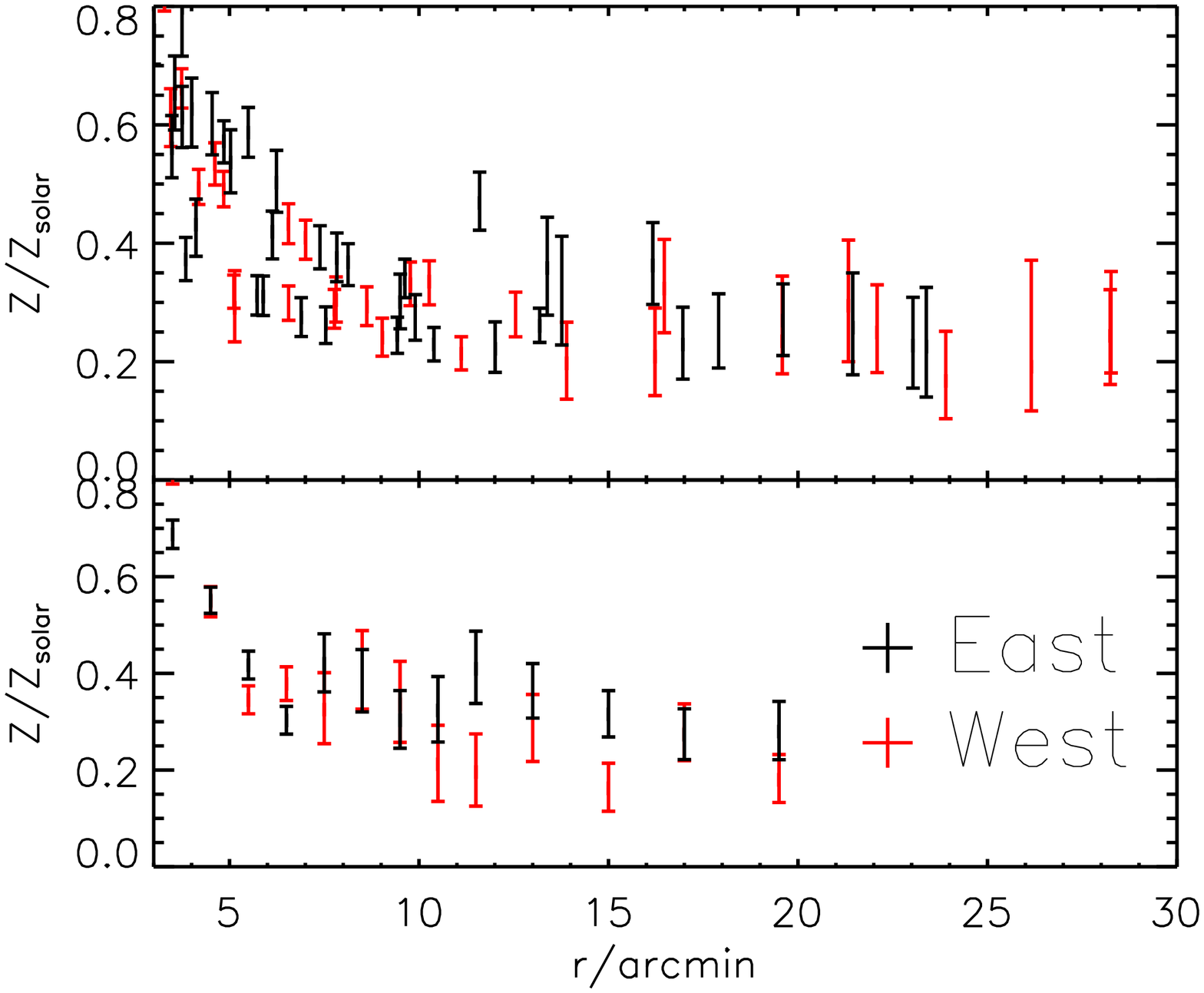,
        width=0.5\linewidth}     
        }   
      \caption{\emph{Top left}:Temperature map using the XMM-MOS data. Annuli are overlaid at radii 5-10-15-20 arcmins to aid comparison with the plots in the lower panels. The temperature map for the central 3 arcmins around NGC 4696 is taken from the Chandra results from \citet{Sanders2006} due to the higher spatial resolution of Chandra. \emph{Bottom left}: Temperature profiles to the east (black) and west (red) corresponding to the temparature map above, showing the higher temperatures to the east. The distance, r, is the distance from the cluster centre of the generators for each region in the Voronoi tesselation. \emph{Top right}: Metallicity map using the XMM-MOS data. Again the central 3 arcmins are from the Fe abundance maps of \citet{Sanders2006} obtained using Chandra. \emph{Bottom right}: Metallicity profiles to the east (black) and west (red). The top shows the values corresponding to the metallicity map, while the bottom shows the values for sectors with equal radial binning to the east and west.}
      \label{TandZmaps}
  \end{center}
\end{figure*}

\begin{figure*}
  \begin{center}
    \leavevmode
      \epsfig{figure=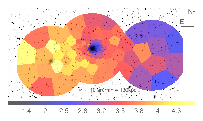,
        width=0.49\linewidth}
         \epsfig{figure=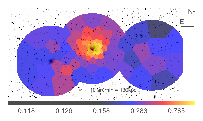,
        width=0.49\linewidth}     
        
      \caption{\emph{Left}:Temperature map overlaid on the DSS optical image.  \emph{Right}: Metallicity map overlaid on the DSS optical image. }
      \label{TandZmaps_withdss}
  \end{center}
\end{figure*}

\section{Background Subtraction and Modelling}
\label{newbackgroundmodelling}

When performing the spectral fitting in \textsc{xspec}, the quiescent particle background (QPB) spectra were subtracted from the extracted MOS spectra. The MOS1 and MOS2 spectra were fitted simultaneously (they were not combined) and the background modelling was performed for each MOS chip individually. 

Two gaussians were used to model the Al K$\alpha$ (1.49 keV) and Cu K$\alpha$ (1.75 keV) instrumental lines with the normalisations allowed to be free parameters. Residual soft proton contamination, which may be present, is modelled as a broken powerlaw as in \citet{Leccardi2008}, with a break at 5keV, a slope of 0.4 below 5keV and a slope of 0.8 above 5.0keV. This soft proton contribution is not folded through the instrumental effective area (the ARF), and this is achieved by using a diagonal unitary response matrix when modelling this component. The low surface brightness, outermost regions of each pointing (outside 10 arcmins for the central pointings, and outside 20 arcmins from NGC 4696 for the east and west offset pointings) were used when fitting for the residual soft proton emission (as recommended in the ESAS cookbook) to ensure that this component is not over or underestimated through confusion with the ICM emission. These normalisations were used to create images of the soft proton contribution to the background using the task \textsc{proton} as described earlier in section \ref{imageproduction} when creating the background subtracted mosaic images.

The cosmic X-ray background (CXB) from unresolved point sources was modelled as an absorbed powerlaw of index 1.46. The normalisation of this powerlaw was calculated using the XMM-ESAS task \textsc{point\_source}, which calculates the unresolved level of emission following the removal of point sources down to the uniform threshold flux of 1$\times$10$^{-14}$ erg cm$^{-2}$ s$^{-1}$ in the 0.7-7.0 keV band. We used the functional form for the logN-logS cumulative source number counts distribution found in \citet{Hasinger2005}. This yields an unresolved CXB powerlaw component with an \textsc{xspec} normalisation of 5.0$\times$10$^{-7}$ per arcmin$^{2}$.

The soft foreground was modelled using an absorbed \textsc{apec} \citep{Smith2001} component at kT=0.22 keV to represent the galactic halo, added to an unabsorbed \textsc{apec} component at kT=0.12keV to represent the Local Hot Bubble (LHB) or heliosphere. Both of these components have their metallicity fixed at 1 $Z_{\odot}$ and their redshift fixed at 0. The temperatures and normalisations of these components were determined by fitting to RASS data for a background region well outside the virial radius of the cluster, between 1.8-2.5 degrees (as done in \citealt{Walker2013_Centaurus}), and are shown in table \ref{GALvariations}. The RASS data were obtained from the X-ray background tool at http://heasarc.gsfc.nasa.gov/cgi-bin/Tools/xraybg/xraybg.pl. An example of the spectral fitting and the background modelling is shown in Fig. \ref{spectralfitting}, showing all of the components involved.

\section{Spectral Analysis}
\label{analysis}

Temperature and metallicity maps were produced and are shown in Fig. \ref{TandZmaps} in the top two panels. We do not explore the central 3 arcmins around NGC 4696 with the \emph{XMM} data because this region has already been explored in detail with deeper and higher spatial resolution Chandra observations (\citealt{Fabian2005}, \citealt{Sanders2006}). The Chandra temperature and metallicity maps from \citet{Sanders2006} for the central 3 arcmins are shown in the respective maps, showing the displacement of the coldest high metallicity gas to the west of NGC 4969. 

Each region was fit with a single temperature absorbed \textsc{apec} component with the metallicity, abundance and normalisation allowed to be free. Spectral fits were performed using the extended C-statistic. The redshift was fixed to 0.0104, and the column density was fixed to the LAB survey \citep{LABsurvey} value of 8.3 $\times$ 10$^{20}$ cm$^{-2}$. We use a standard $\Lambda$CDM cosmology with $H_{0}=70$  km s$^{-1}$ Mpc$^{-1}$, $\Omega_{M}=0.3$, $\Omega_{\Lambda}$=0.7.  Spectral analysis was performed in \textsc{xspec} 12.8, and fits were performed in the 0.7-7.0 keV band. We use the abundance tables of \citet{Anders1989} so as to be consistent with the results of \citet{Sanders2006}. All errors unless
otherwise stated are at the 1 $\sigma$ level.

\subsection{Temperature and metallicity maps}

In the lower two panels in Fig. \ref{TandZmaps} the temperature and metallicity profiles are shown corresponding to the regions in the maps in the upper panels, allowing the errors on the profiles to displayed and allowing a comparison of the profiles to the east and west. The temperature is systematically higher to the east by on average around 0.6 keV in the radial range 5-22 arcmins, in good agreement with the observed temperature excess around Cen 45 observed in \citet{Churazov1999} with \emph{ASCA} data. 

The metallicity map suggests that there is a metal abundance excess to the east associated with Cen 45. This becomes clearer when we compare the east and west in sectors with equal radial binning in the bottom part of the bottom right hand panel of Fig. \ref{TandZmaps}, where we see an excess to the east between 11-12 arcmins. When we extract and fit spectra to the whole regions between 10-15 arcmins from the core to the east and west, we find that the abundance is significantly higher to the east (0.36$_{-0.05}^{+0.05}$ Z$_{\odot}$) than to the west (0.21$_{-0.05}^{+0.05}$ Z$_{\odot}$). This indicates that Cen 45 has retained its gas and metals during the interaction with the main Centaurus cluster. 

\begin{figure}
  \begin{center}
    \leavevmode
    \vbox{
         \epsfig{figure=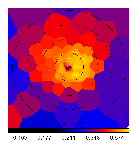,
        width=0.9\linewidth} 
      \epsfig{figure=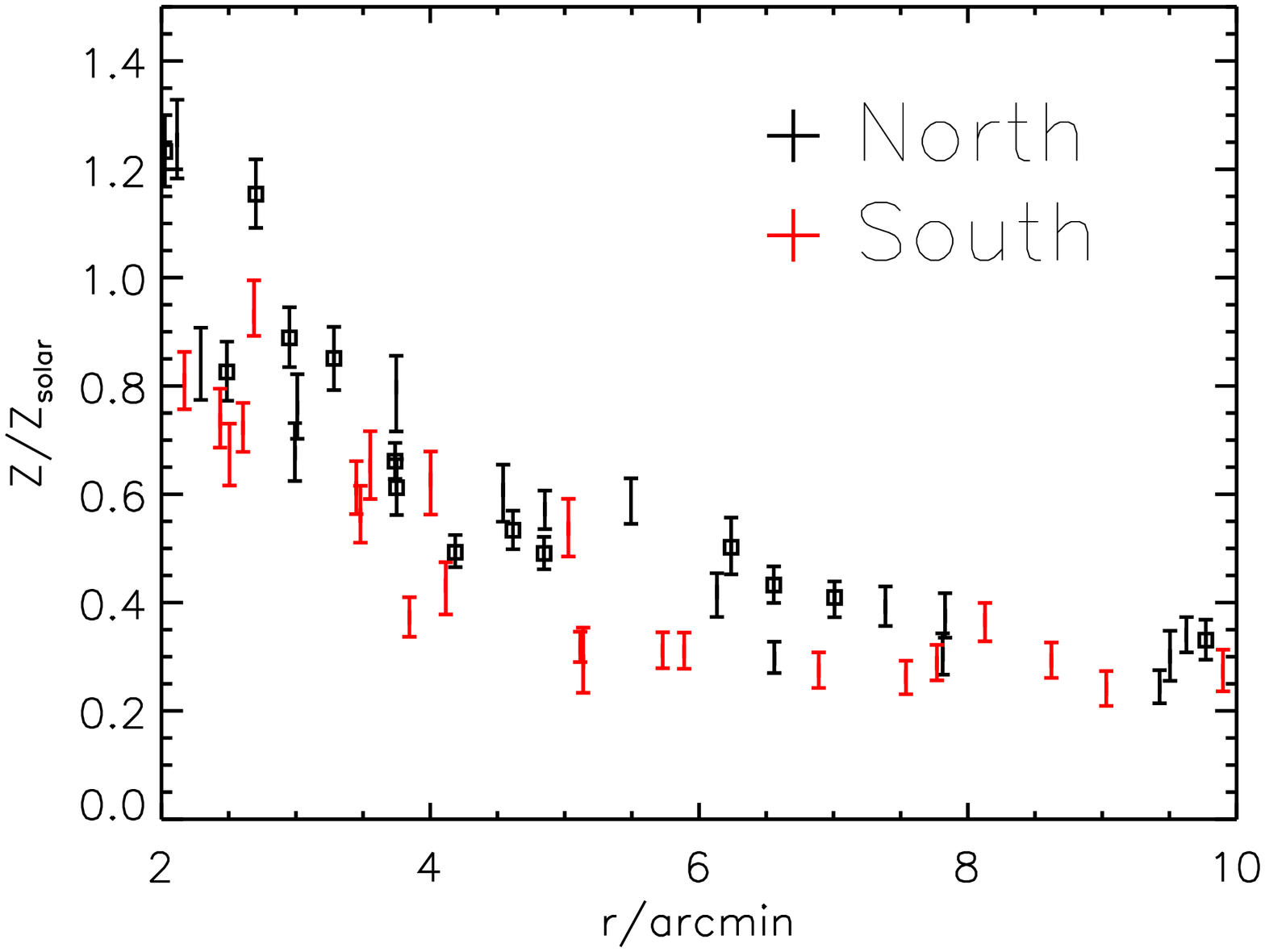,
        width=\linewidth}  
        }
          \caption{The top panel shows a zoom in of the central 8 arcmins of the metallicity map shown in Fig. \ref{TandZmaps}, with annuli shown at 2, 4, 6 and 8 arcmins. The bottom panel compares the metallicity profiles to the north and south around the Centaurus cluster core, showing the excess in metallicity to the north between 4-8 arcmins. The points marked with a square are associated with the filamentary structure to the north west found later and shown in Fig. \ref{Texcess_withdss}.}
      \label{Z_north_south}
  \end{center}
\end{figure}

The metallicity distribution is also clearly higher to the north of the core of Centaurus than to the south. As can be seen in Fig. \ref{Z_north_south} there is a clear excess to the north in the region 4-8 arcmins. This excess in metallicity to the north may be associated with the filamentary structure to the north west we detect later in section \ref{filament}, which was also detected in the \emph{ROSAT} analysis of \citet{Churazov1999}, and which appears to connect NGC 4696 to NGC 4696B. We will discuss this further in section \ref{filament}.
 
The temperature and metallicity maps are overlaid on the DSS image of the galaxies in Fig. \ref{TandZmaps_withdss} to show how the ICM properties correlate with the galaxy distribution. We see a clear excess in temperature to the west, associated with Cen 45, where the ICM is systematically at least 0.6 keV hotter than the equivalent regions to the east. The hotter region extends out to a radius of around 10 arcmin ($\sim$ 130 kpc) from NGC 4709. 

\subsection{Entropy and pressure}
\label{entropyandpressure}
To obtain the entropy, $K=kT/n_{e}^{2/3}$, and the pressure, $P=n_{e}kT$ direct from the spectral fitting results, we need to covert the observed \textsc{apec} normalisations into densities, which requires us to estimate volumes for each region. To do this, we follow the the approach of \citet{Henry2004} (as used in \citet{Simionescu2007}). The electron density can be related to the \textsc{apec} normalisation and the volume, V of each region by,

\begin{equation}
n_{e} = 2.18 \times 10^{-5} h_{70}^{-1/2} \rm{cm^{-3}} \sqrt{\frac{norm_{apec}}{V (\rm{Mpc}^3) }}D_{A} \rm{(Mpc)} (1+z)
\label{eq:volumes}
\end{equation}

The volume of each region can then be estimated as $V\approx(4/3)D_A^3\Omega(\theta_{out}^2-\theta_{in}^2)^{1/2}$, where $\Omega$ is the solid area of the region, and $\theta_{out}$ and $\theta_{in}$ are the outer and inner angular distances of the region from the centre. 

\begin{figure}
  \begin{center}
    \leavevmode
    
    \vbox{
      \epsfig{figure=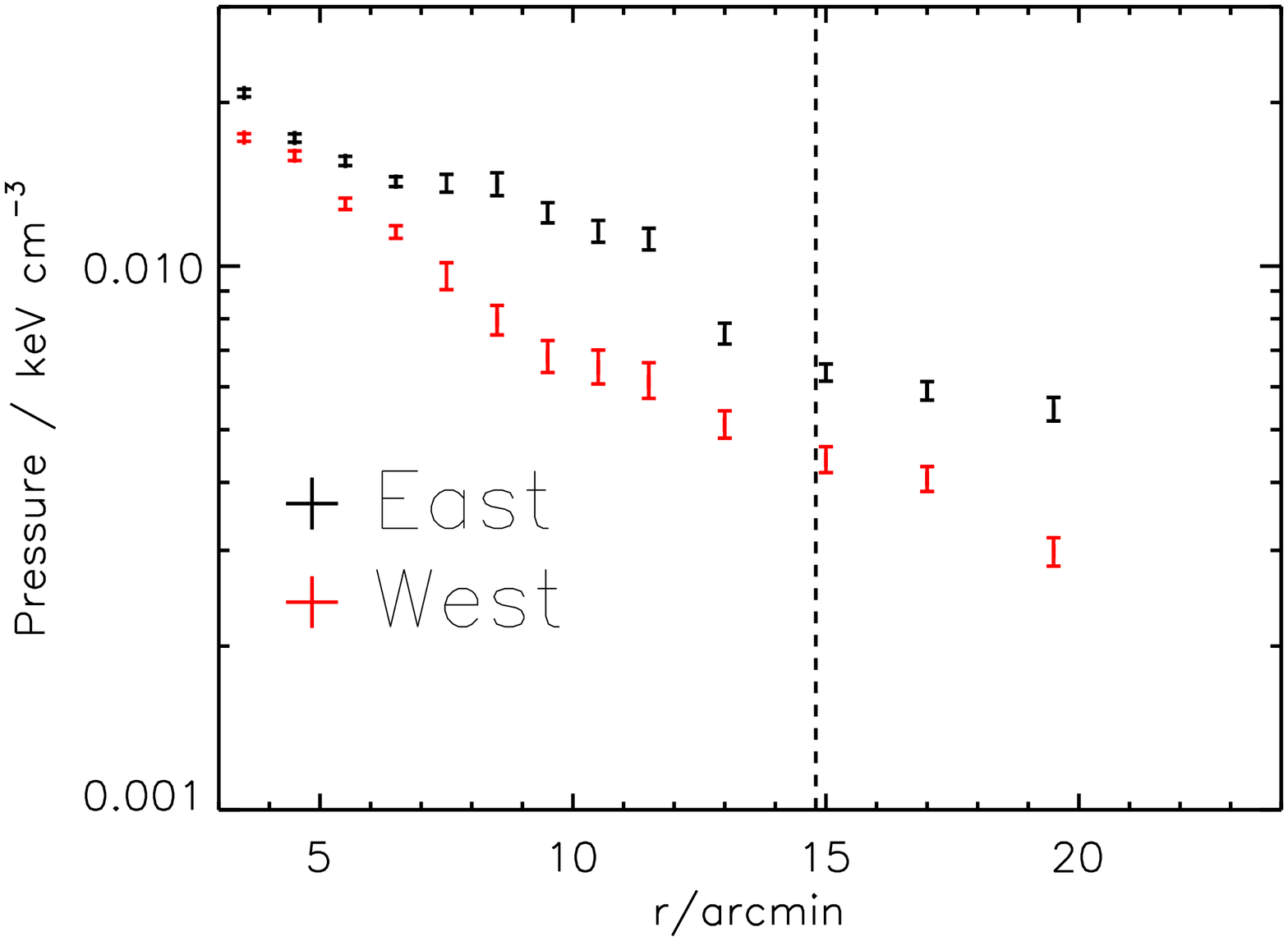,
        width=0.95\linewidth}
      \epsfig{figure=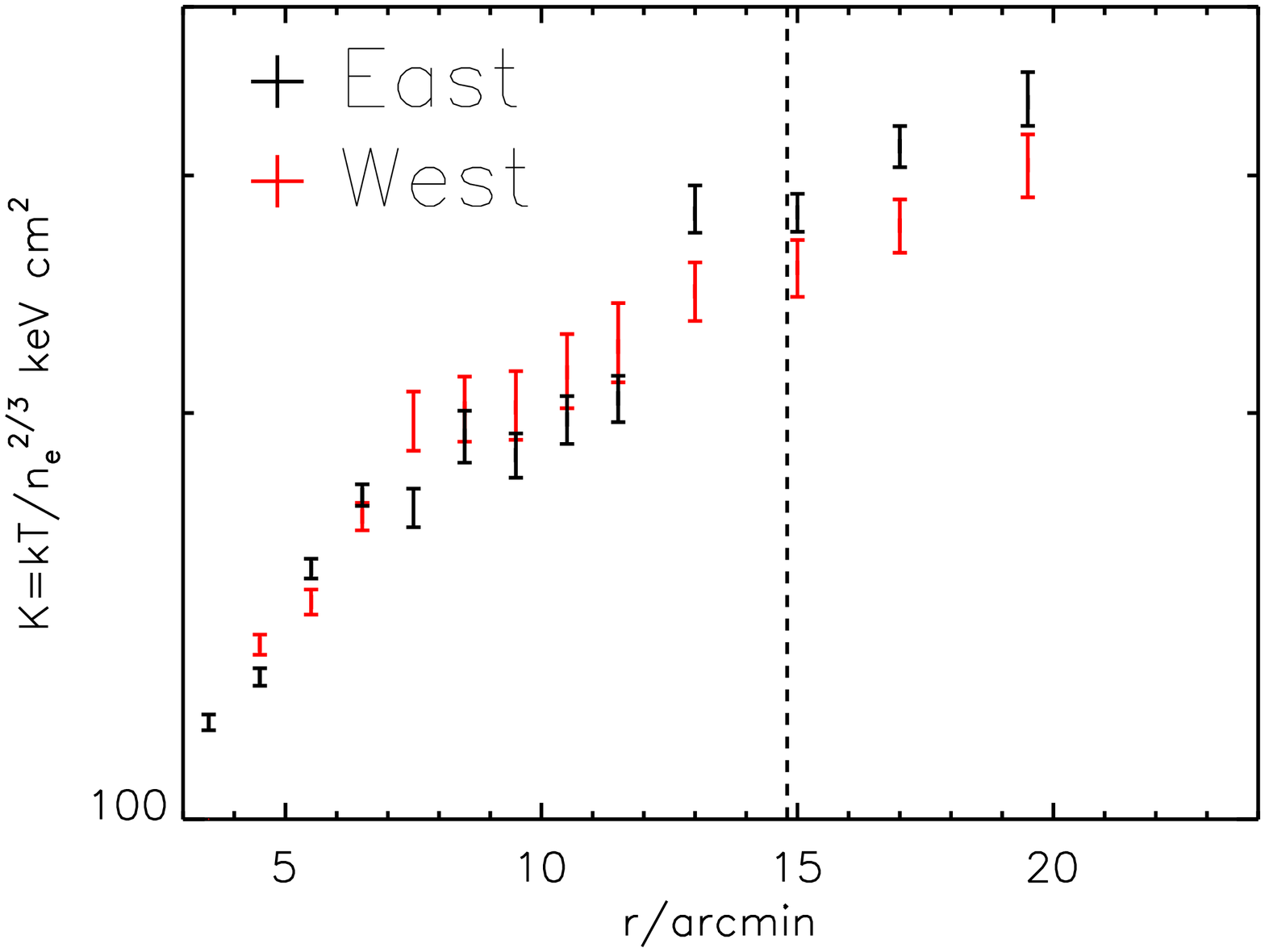,
        width=0.95\linewidth}   
   
        }

      \caption{\emph{Top}:Projected pressure profiles to the east (black) and the west (red) along sectors with equal radial binning. \emph{Bottom}:Projected entropy profiles to the east (black) and the west (red). In both plots the vertical dashed line shows the distance of the central galaxy of Cen 45 (NGC 4709) from the core of the main Centaurus cluster. }
      \label{pressureandentropy}
  \end{center}
\end{figure}

To compare the entropy and pressure to the east and west, we extracted spectra in sectors to the east and west. The pressure profiles derived from the spectral fits are shown in the top panel of Fig. \ref{pressureandentropy}. We see a clear excess in pressure to the east in the direction of the Cen 45 merger, in agreement with the pressure map derived from the hard band image in Fig. \ref{pressuremap}. 

Outside 12 arcmins the entropy to the east is systematically higher than that to the west as we would expect for shock heating. Within 12 arcmins, the temperature and density are higher to the east in such a way that the entropy is similar to the entropy to the west, but the pressure is higher to the east. This may be because the ICM to the west has already undergone similar merging activity in the past which has raised it to a similar adiabat.

\subsection{Cen 45}

\subsubsection{Excess emission around Cen 45}

\begin{figure}
  \begin{center}
    \leavevmode
    \vbox{
    \epsfig{figure=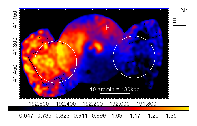,
        width=\linewidth}
      \epsfig{figure=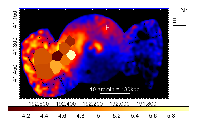,
        width=\linewidth}
      \epsfig{figure=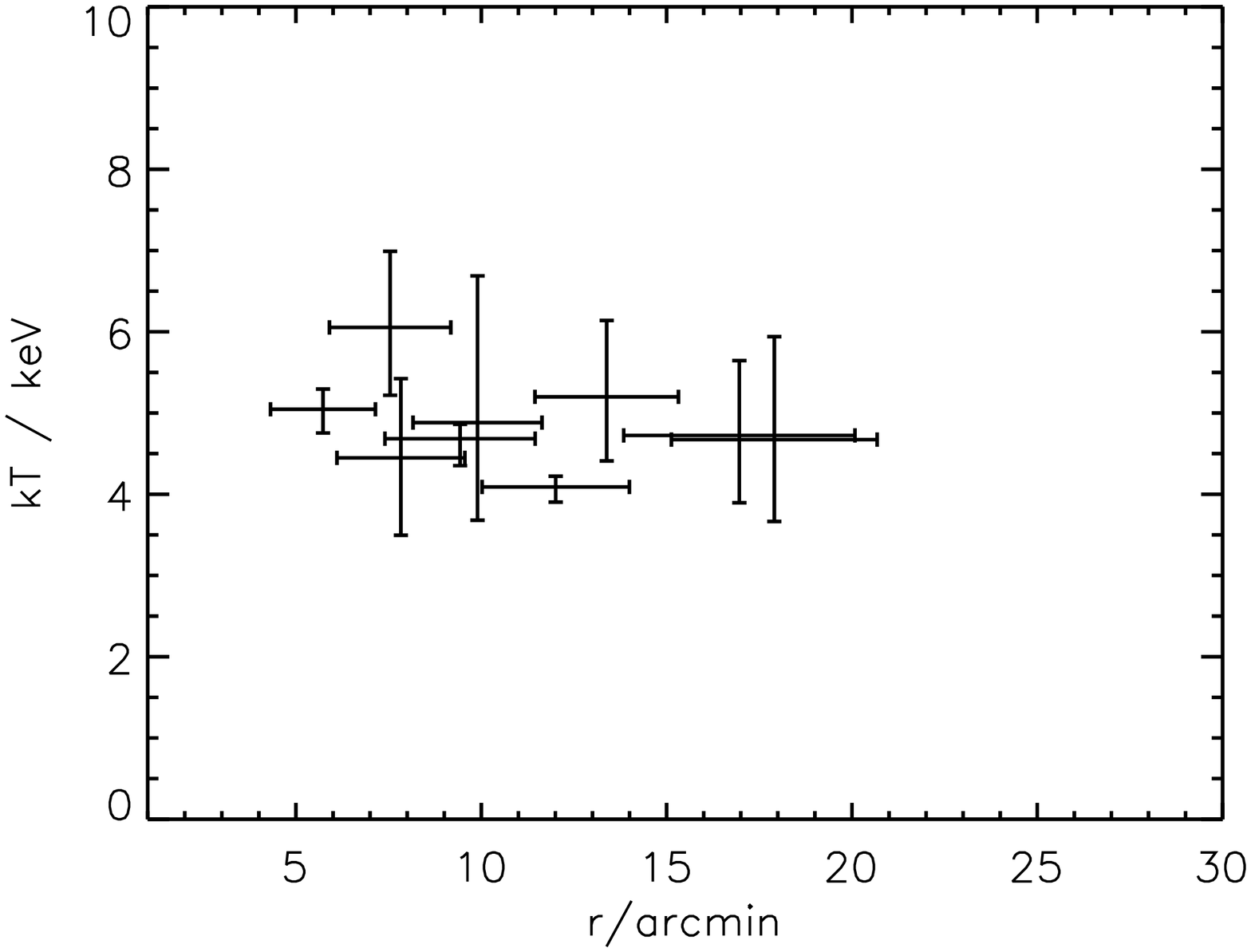,
        width=\linewidth}   
   
        }

      \caption{\emph{Top}:Residual image from dividing the 0.7-2.0 keV band image by the azimuthal average, which has been smoothed with a Gaussian kernel of radius 40 arcseconds. The dark straight lines are from the chip gaps between the MOS detector chips. The letter F marks the excess in emission to the north west which agrees with the filamentary structure detected in the ROSAT analysis of \citet{Churazov1999} between NGC 4696 and NGC 4696B. The circle to the East encompasses a large fraction of the emission associated with Cen45, and this is the region examined in Fig. \ref{2cmp_spec}. The equal sized circle to the west is at the same distance from the core as the western circle. \emph{Middle}:Same as the top panel but with the temperature map of the excess emission overlaid (the colour bar is for the temperatures). \emph{Bottom}:Temperature profile of the excess emission.  }
      \label{Texcess_withdss}
  \end{center}
\end{figure}

\citet{Churazov1999} divided the \emph{ROSAT} PSPC image by the azimuthal average and found a peak in the residuals coinciding with Cen 45 due to excess emission. Dividing the 0.7-2.0 keV \emph{XMM} mosaic image by the azimuthal average (Fig. \ref{Texcess_withdss}, top) also shows a significant excess in the X-ray emission above the azimuthal average coinciding with Cen 45 to the east (though as the \emph{XMM} mosaic does not cover the full azimuth this only compares the east and west offset pointings). 

If we are seeing the superposition of the ICM of Cen 45 with the Centaurus cluster, we can explore the gas properties around Cen 45 by including the emission from the main Centaurus cluster in the background model. To do this, in the regions of surface brightness excess to the east we redo the spectral fits using a 2 thermal (\textsc{apec}) component model, with one \textsc{apec} component fixed to the temperature, normalisation and metallicity at the equivalent distance from the NGC4696 from the western pointing. 

The temperature map of the excess emission is shown in Fig. \ref{Texcess_withdss} (middle panel), overlaid on the surface brightness residual image. The emission has a temperature of around 5 keV (between 4 keV and 6 keV), roughly 2 keV higher that of the ICM of the main Centaurus cluster in the western pointing. The temperature profile of the excess emission is plotted in the bottom panel of Fig. \ref{Texcess_withdss}. 

\begin{figure}
  \begin{center}
    \leavevmode
    \epsfig{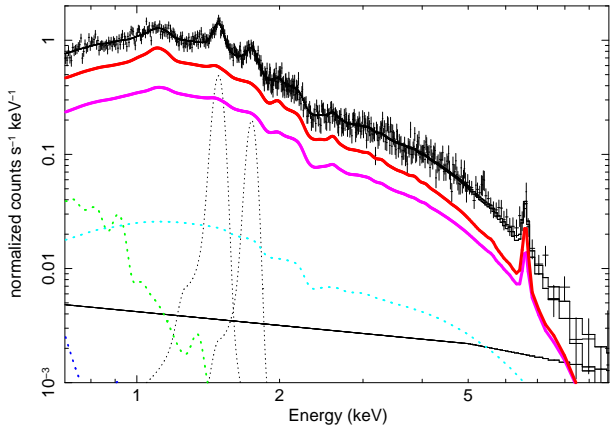}

      \caption{Best fit two-temperature model for the spectrum of the region around Cen 45 (the white circle to the east in Fig. \ref{Texcess_withdss}). The solid red line is the apec component whose normalisation, temperature and metallicity are fixed to the best fit single temperature model of the region at equal distance from the centre on the opposite side of the cluster to the excess emission (the white circle to the west in Fig. \ref{Texcess_withdss}). The solid pink line is the apec component modelling the excess emission associated with Cen 45 to the east. The background components are the same as described in Fig. \ref{spectralfitting}.  }
            \label{2cmp_spec}
  \end{center}
\end{figure}

To show an example of the spectral fitting, we show in Fig. \ref{2cmp_spec} the best fit spectrum of the entire excess region using the two temperature model. The spectrum is for the entire excess region shown by the circle to the east in the top panel of Fig. \ref{Texcess_withdss}. In the two temperature model, one of the \textsc{apec} components has its temperature, abundance and normalisation fixed to that found in an equal sized circle at the same distance from NGC 4696 to the west which is also shown in the top panel of Fig. \ref{Texcess_withdss}, and this is the red solid line in Fig. \ref{2cmp_spec}. The best fit two temperature model parameters are shown in table \ref{2cmp_parameters}.

It has been found in \citet{Mazzotta2004} that when the lowest dominant temperature component in a multitemperature source spectrum is greater than 2-3keV (as it is in our case), it is always possible to fit the spectrum with a statistically acceptable single temperature thermal model. When the eastern region is fitted with a single temperature model (the details of which are also shown in table \ref{2cmp_parameters}), we do indeed obtain an equally good fit as the two temperature model (with a reduced $\chi^{2}$ of 0.93 in both cases). It is therefore not possible to determine from the spectra whether the two temperature model is preferred.

If Cen 45 has retained its own gas, as suggested by the excess in metal abundance near its location, we would expect to see indications of a cool component in Fig. \ref{2cmp_spec}. The absence of such indications may be because the cool component is too faint to be detectable, or another possibility is that the ICM of Cen 45 has been heated to high temperatures due to its interaction with the main Centaurus cluster.

\subsubsection{Thermal energy content of the excess emission around Cen 45}

If we assume that the ICM around Cen 45 is shock heated under the simple Rankine-Hugoniot strong shock jump conditions then we can, as a first approximation, estimate the temperature increase per particle assuming the shock speed to be equal to the velocity difference between Cen 45 and the main Centaurus cluster. The shock speed would therefore be $u_{1}$=1500/$\cos(\theta)$ km s$^{-1}$, where $\theta$ is the angle to the line of sight of the motion of Cen 45. 

\begin{figure}
  \begin{center}
    \leavevmode
    \epsfig{figure=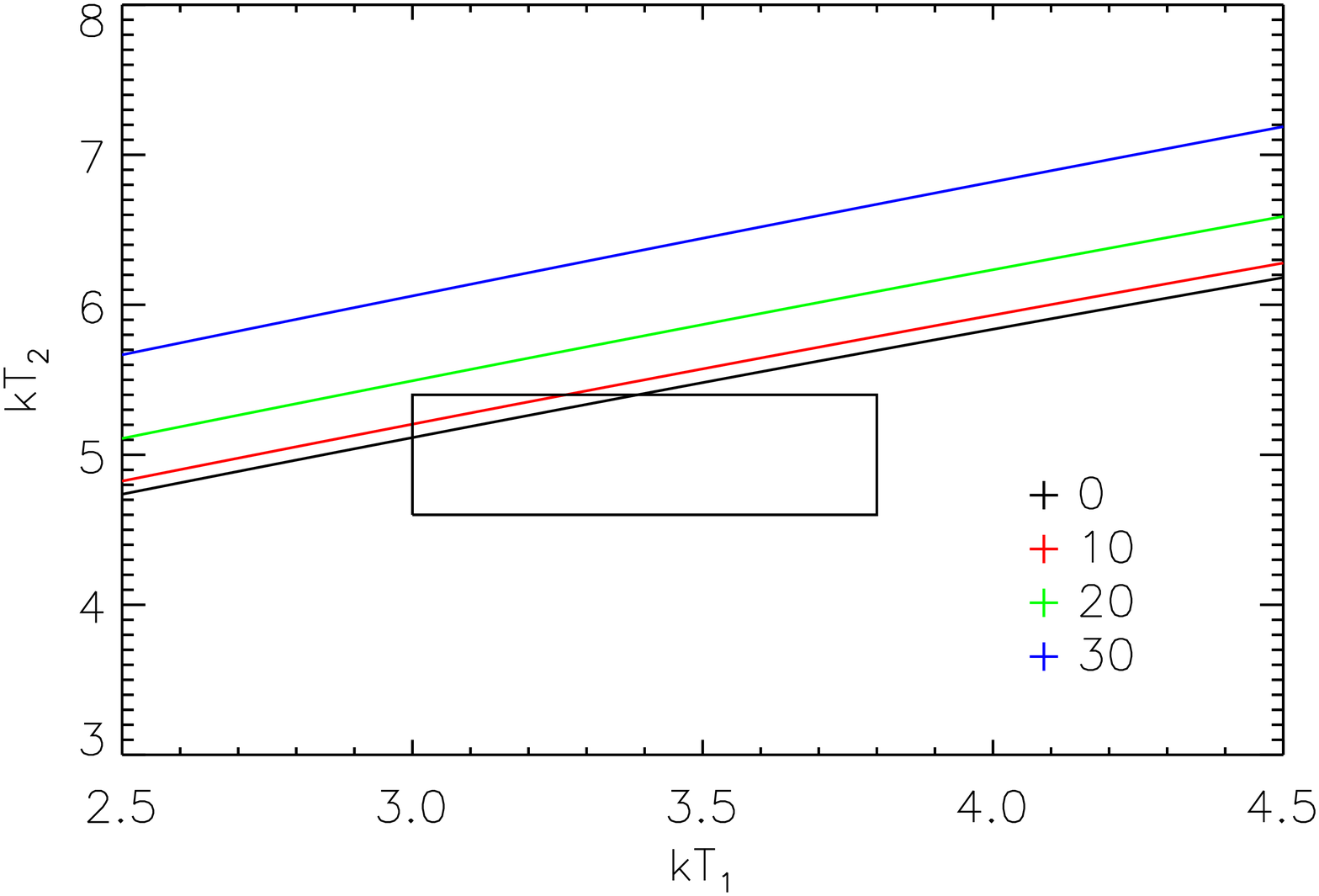,
        width=\linewidth}

      \caption{Post shock temperature (kT$_{2}$) as a function of pre-shock temperature (kT$_{1}$) and the angle to the line of sight of Cen 45's motion (the coloured lines). The black rectangle shows the region encompassing the expected pre-shock temperature based on observations of the undisturbed western side of the cluster, and the observed post-shock temperature range.  }
      \label{Fig:shockheating}
  \end{center}
\end{figure}

In Fig. \ref{Fig:shockheating} we show how the temperature per particle after the shock (kT$_{2}$) depends on the pre-shock temperature per particle upstream of the shock (kT$_{1}$) and the angle to the line of sight ($\theta$) of the motion of Cen 45. The greater the angle to the line of sight, the greater the shock speed, and thus the greater the amount of heating. This is calculated using:

\begin{align}
\frac{T_{2}}{T_{1}} &= \frac{[(\gamma-1)M_{1}^{2} + 2][2 \gamma M_{1}^{2} - (\gamma -1 )]}{(\gamma+1)^{2}M_{1}^{2}}
\label{eqn:shockheatinggeneralT}
\end{align}
\begin{align}
M_{1} &= \left( \frac{mu_{ \mathrm{shock} }^{2}}{\gamma kT_{1}}\right)^{1/2} 
\label{eqn:shockheatinggeneralMach}
\end{align}
\begin{align}
u_{\mathrm{shock}} = \frac{1500}{\cos(\theta)} \rm{kms^{-1}}
\label{eqn:shockspeed}
\end{align}
where $m$ is the mass per particle (assumed to be 0.61m$_{H}$), and the adiabatic index $\gamma$=5/3.

We can calculate an estimate of the pre-shock, upstream temperature by using the temperature for the undisturbed western region (the western circle shown in Fig. \ref{Texcess_withdss}), which is kT$_{1}$=3.4$^{+0.4}_{-0.4}$ keV. We can then take the post-shock temperature to be the temperature of the high temperature component in the two temperature model fit to the excess around Cen 45 (the eastern circle in Fig. \ref{Texcess_withdss}), which is kT$_{2}$=5.0$^{+0.4}_{-0.4}$ keV.
   
In Fig. \ref{Fig:shockheating}, the black rectangle encompasses the observed post-shock temperature range and the pre-shock temperature range we have just calculated. We see that the shock heating scenario is consistent with the observed temperature difference, and that an angle to the line of sight greater than 20 degrees appears unlikely. The allowed range of the 3D velocities for the interaction with the subcluster is 1300-1600 km s$^{-1}$. 

\begin{table}
  \begin{center}
  \caption{Best fit parameters to the eastern region around Cen 45 (the eastern white circle in Fig. \ref{Texcess_withdss}) using two temperature and single temperature models. In the two temperature model, component 2 is fixed to the best fit parameters for a region on the opposite side of the cluster (the white circle to the west in Fig. \ref{Texcess_withdss}) of equal size and equal distance from the cluster centre. Normalisations are in xspec units per arcmin$^{2}$.}
  \label{2cmp_parameters}
  
    \leavevmode
    \begin{tabular}{lll} \hline \hline
   & Two T model & One T model \\ \hline
   $kT_{1}$/keV & 5.0$^{+0.4}_{-0.4}$ & 4.0$^{+0.2}_{-0.2}$ \\
   $Z_{1}/Z_{\odot}$ & 0.27$^{+0.09}_{-0.08}$ & 0.29$^{+0.04}_{-0.04}$\\
   norm$_{1}$ & 4.8$^{+0.2}_{-0.2}$ $\times$ 10$^{-5}$ & 14.0$^{+0.2}_{-0.2}$ $\times$ 10$^{-5}$ \\ \hline
   $kT_{2}$/keV & 3.4 & - \\
   $Z_{2}/Z_{\odot}$ &0.24 & - \\
   norm$_{2}$ & 9.2 $\times$ 10$^{-5}$ & - \\ \hline
      $\chi^{2}$/d.o.f. & 532.26/572 & 533.61/572 \\ \hline

    \end{tabular}
  \end{center}
\end{table}

We conclude that the simplest model of strong shock heating provides an adequate explanation for the high temperature of the gas surrounding Cen 45, but that due to the limitations of the spectral fitting it is not possible to determine statistically whether the two temperature scenario is favoured just from the X-ray spectra. 

\section{Filament extending to NGC 4696B}
\label{filament}
When dividing by the azimuthal average there is also a clear excess in X-ray emission connecting NGC 4696 to NGC 4696B, indicated by the letter F (for filament) in the top panel of Fig. \ref{Texcess_withdss}. This filamentary structure was observed with much lower significance (2-3 $\sigma$) in the \emph{ROSAT} analysis of \citet{Churazov1999}. Unfortunately, due to the limited azimuthal coverage of the \emph{XMM-Newton} observations, the full extent of this filamentary structure cannot be fully observed. In \citet{Churazov1999} it is suggested that this filament could be the result of cold gas which has been ram pressure stripped from NGC 4696B. 

The metallicity map shown in Fig. \ref{TandZmaps} (right panel) indicates that there is an abundance enhancement to the north of NGC 4696, which may be associated with this filament. If so, this would add support to the idea that the filament originates from gas stripped from NGC 4696B. However the lack of detection of an excess in metallicity in the other parts of the filament closer to NGC 4696B may suggest that the northern metallicity excess is due to the past sloshing of metals originating from NGC 4696. The Chandra results of \citet{Fabian2005} and \citet{Sanders2006} found the high metallicity gas within the central 40 kpc to by displaced to the west of NGC 4696. It is therefore possible that in the past, previous merging activity could have caused a northern displacement of the core leading to the observed asymmetric distribution of metals out to 100kpc from the BCG. 

\section{Summary}

We confirm the presence of a significant excess in temperature to the east of the Centaurus cluster core associated with the subgroup Cen 45, as originally found using \emph{ASCA} data in \citet{Churazov1999}, \citet{Furusho2001} and \citet{Dupke2001}. Our results indicate that minor mergers are able to significantly heat the ICM of galaxy clusters in the region surrounding the core without destroying the cool core. A simple shock heating scenario appears to be able to provide a reasonable explanation for the observed high temperatures surrounding Cen 45. The metallicity around Cen 45 is significantly higher than that in the western direction at the equivalent distance from the centre of the main Centaurus cluster, suggesting that Cen 45 has managed to retain its gas as it has interacted with the main Centaurus cluster. 

We confirm the detection of a filamentary structure extending between NGC 4696 and NGC 4696B which was detected with lower significance in the \emph{ROSAT} analysis of \citet{Churazov1999}.

We have probed the metallicity structure around the BCG of Centaurus (NGC 4696) out to at least 150 kpc in all directions, extending the existing \emph{Chandra} observations of the central 40 kpc regions. The metallicity to the north of NGC 4696 is higher than to the south, which may indicate the asymmetric movement of metals through the sloshing motion of the core in the past. Alternatively, this higher metallicity may be associated with the filamentary structure extending between NGC 4696 and NGC 4696B. 

The pressure is higher to the east in the direction of Cen 45, and outside 12 arcmins the entropy profile is systematically higher to the east than to the west, as expected from shock heating.

\label{summary}

\section*{Acknowledgements}

SAW is supported by STFC, and ACF thanks the Royal Society. This
work is based on observations obtained with \emph{XMM-Newton}, an ESA
science mission.

\bibliographystyle{mn2e}
\bibliography{XMM_Centaurus_paper}

\appendix
\section[]{}
\label{sec:appendix}

\subsection{Checking with PN data}
\label{PN_data}

To check our MOS results we extract spectra from the PN data for each region of the Voronoi tessellation, using the XMM-ESAS tools. The pn data were filtered using \textsc{epchain} and \textsc{pn-filter}, and we checked for any residual soft proton flares by examining the hard band (10-14 keV) light curves and excluding periods where the rate was more than 3 $\sigma$ from the mean. Point sources were removed using \textsc{cheese} down to the same threshold flux as for the MOS data (1 $\times$ 10$^{-14}$ erg cm$^{-2}$ s$^{-1}$). Spectra, RMFs and ARFS for each region in the tessellation were then extracted from the pn data using \textsc{pn-spectra}, and quiescent particle background spectra for each region were obtained using \textsc{pn-back}. 

The spectra were then fit in the same way as for the MOS data by subtracting the quiescent particle background and using the same background model, with the only difference being that 6 instrumental lines need to be modelled for the pn detector using gaussians at energies 1.49, 7.49, 7.11, 8.05, 8.62 and 8.90 keV, whose normalisations were free parameters. 

\begin{figure*}
  \begin{center}
    \leavevmode
    \hbox{
    \epsfig{figure=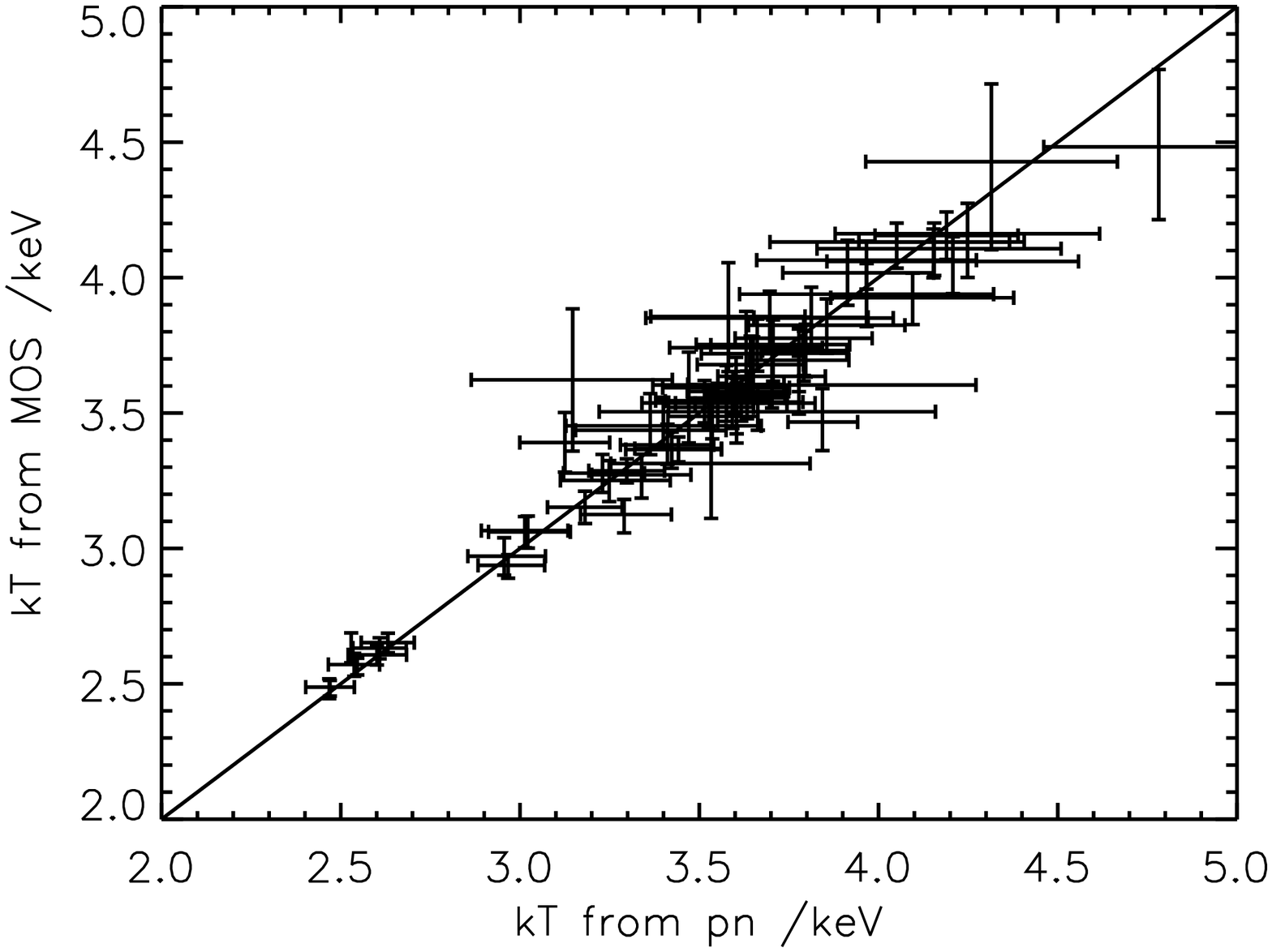,
        width=0.45\linewidth, angle=0}
    \epsfig{figure=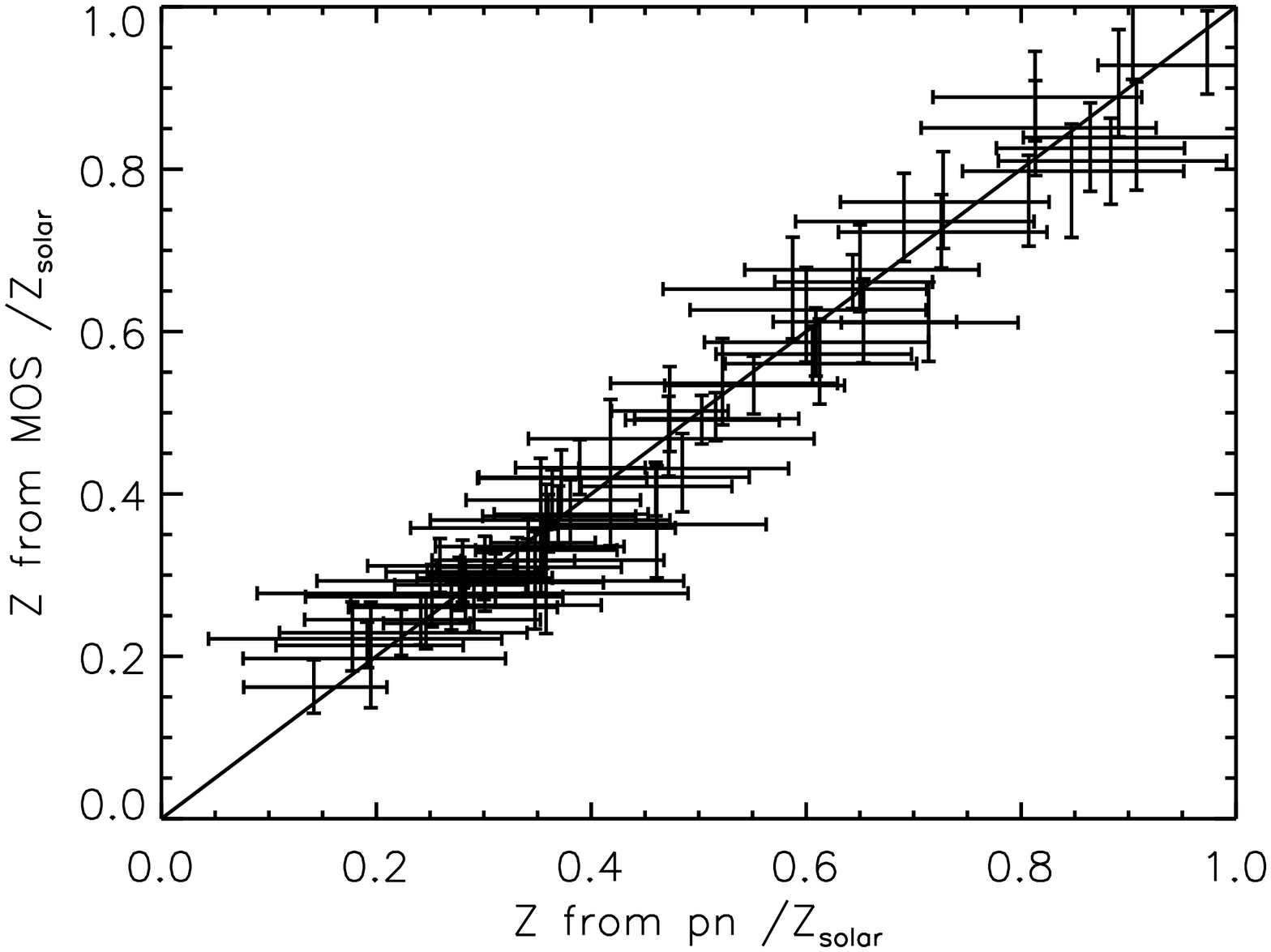,
        width=0.45\linewidth, angle=0}
}

\caption{Plotting the temperatures (left panel) and metallicities (right panel) found with the MOS data against those found with the pn data for the regions in the tessellation.}
   \label{pn_mos_compare}
  \end{center}
\end{figure*}

In Fig. \ref{pn_mos_compare} we show the temperatures and metallicities for each region obtained with MOS plotted against the values obtained by fitting to the pn data; in all cases there is good agreement between the MOS and pn results. There is expected to be a systematic difference between the temperatures from the MOS and pn detectors, as found in \citealt{Nevalainen2010}, and the good agreement we is observe is likely due to the statistical errors of the fits dominating over the systematic differences.

We also repeat the analysis of the eastern and western circles shown in Fig. \ref{Texcess_withdss} using the pn data. The sensitivity to multi temperature spectra of the pn detector is different to that of the MOS, and this allows us to check our result for the two temperature model. The results for fitting the pn data in the eastern region around Cen 45 with a two temperature model appear consistent with the MOS results, and these are tabulated in table \ref{pn:2cmp_parameters}.

\begin{table}
  \begin{center}
  \caption{The same as table \ref{2cmp_parameters}, but this time using the pn data for the eastern and western circles shown in Fig. \ref{Texcess_withdss}. }
  \label{pn:2cmp_parameters}
  
    \leavevmode
    \begin{tabular}{lll} \hline \hline
   & Two T model & One T model \\ \hline
   $kT_{1}$/keV & 4.8$^{+0.4}_{-0.4}$ & 3.8$^{+0.2}_{-0.2}$ \\
   $Z_{1}/Z_{\odot}$ & 0.31$^{+0.09}_{-0.08}$ & 0.29$^{+0.05}_{-0.05}$\\
   norm$_{1}$ & 4.8$^{+0.2}_{-0.2}$ $\times$ 10$^{-5}$ & 14.0$^{+0.2}_{-0.2}$ $\times$ 10$^{-5}$ \\ \hline
   $kT_{2}$/keV & 3.2 & - \\
   $Z_{2}/Z_{\odot}$ &0.23 & - \\
   norm$_{2}$ & 9.2 $\times$ 10$^{-5}$ & - \\ \hline
      $\chi^{2}$/d.o.f. & 1073/976 & 1077/976 \\ \hline

    \end{tabular}
  \end{center}
\end{table}

\begin{figure*}
  \begin{center}
    \leavevmode
    \epsfig{figure=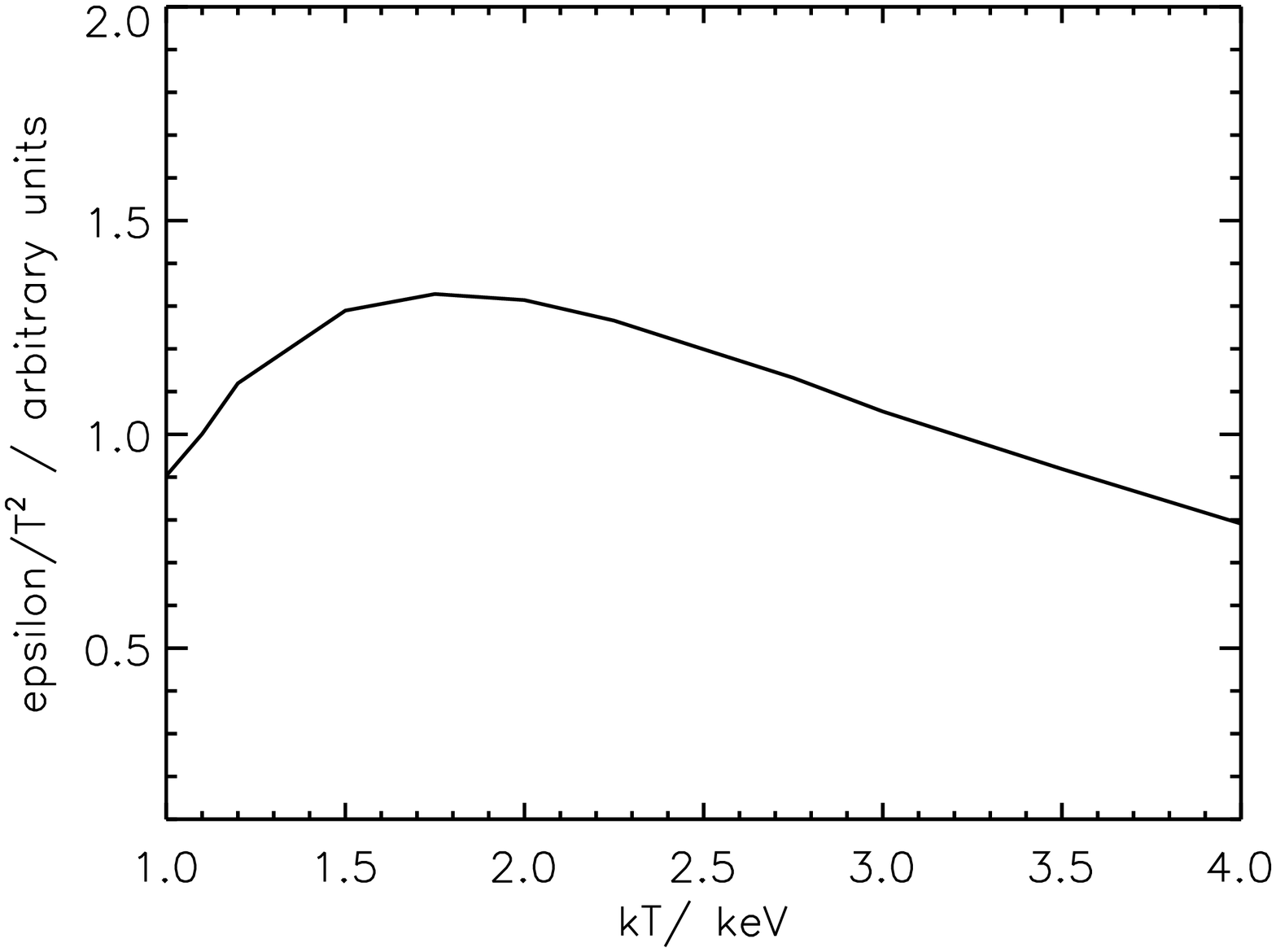,
        width=\linewidth, angle=0}

      \caption{Plot of the temperature dependence of $\epsilon(T)/T^2$ for the hard band (3.5-7.5keV) as observed with the \emph{XMM-Newton} MOS detectors, showing the small level of temperature dependence. This allows the 3.5-7.5keV image to be used as a map of the squared pressure integrated along the line of sight.}
      \label{hardbandindependence}
  \end{center}
\end{figure*}

\begin{figure*}
  \begin{center}
    \leavevmode
    \epsfig{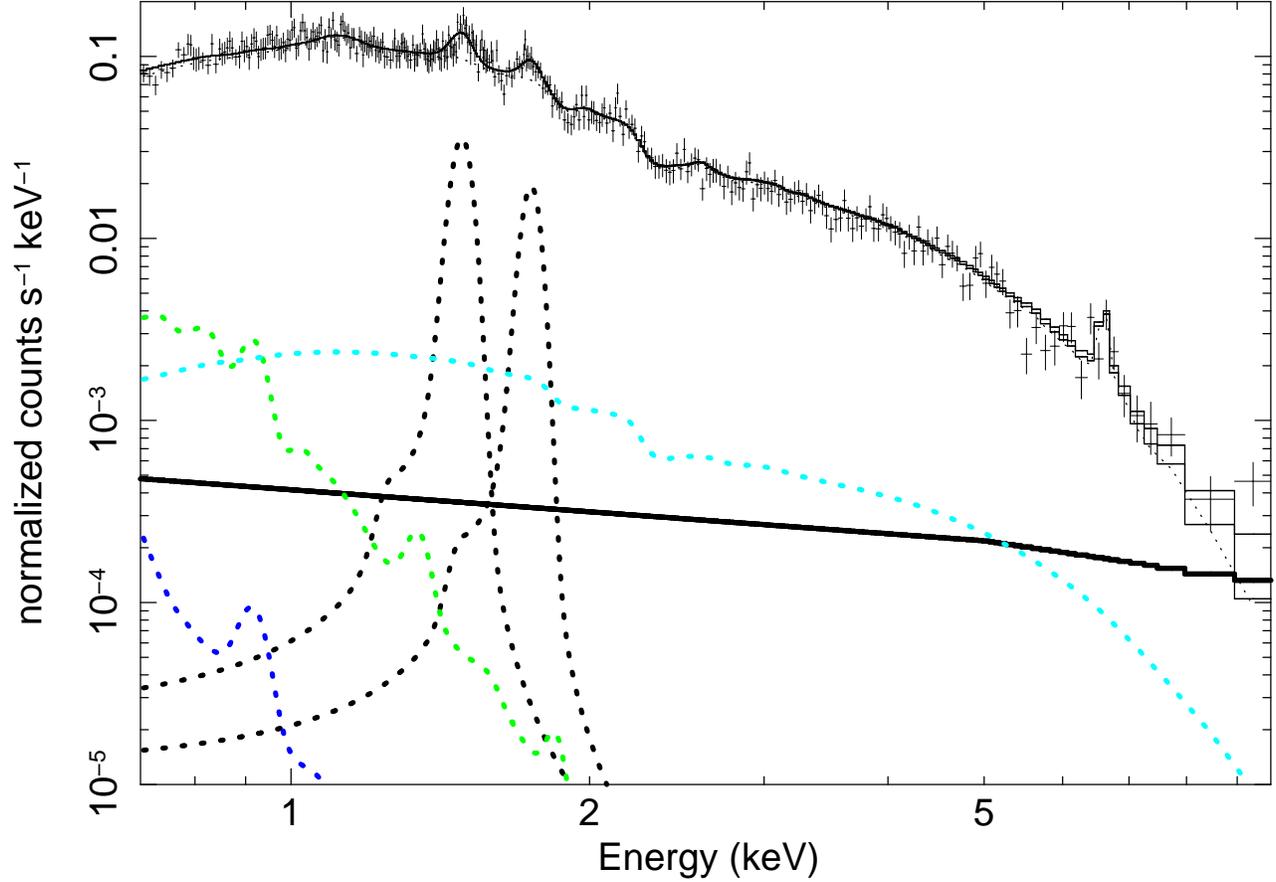}

      \caption{An example of the spectral fitting performed, shown for one of the regions in the adaptively binned Voronoi tesselation (Fig. \ref{TandZmaps}) which contains 20000 counts, and this particular region is 8 arcmins from the centre of the main Centaurus cluster. The black datapoints are MOS1 data, from which the particle background obtained using \textsc{mos-back} has been subtracted. For each region we fit the MOS1 and MOS2 data simultaneously, but here we show only the MOS1 data to improve the figure clarity. The soft foreground is modelled as an absorbed \textsc{apec} component at kT=0.22 keV to represent the galactic halo (green dashed line), added to an unabsorbed \textsc{apec} component at kT=0.12keV (blue dashed line) to represent the Local Hot Bubble (LHB) or heliosphere. These foreground components are constrained by fitting to RASS data in a background ring around the Centaurus cluster as described in the main text. The CXB component for the \emph{XMM} data is modelled as a powerlaw of index 1.46 and is shown as the cyan line. The unresolved CXB level following point source removal is calculated using the XMM-ESAS task \textsc{point-source}. The black dashed lines show the Gaussians used to model the the Al K$\alpha$ (1.49 keV) and Cu K$\alpha$ (1.75 keV) instrumental lines. The excess cluster emission is modelled as an absorbed \textsc{apec} component.  Residual soft proton emission was also modelled as a broken powerlaw component, shown as the solid black line.}
      \label{spectralfitting}
  \end{center}
\end{figure*}



\end{document}






















%% file: XMM_Centaurus_paper.bbl
\begin{thebibliography}{}

\bibitem[\protect\citeauthoryear{{Anders} \& {Grevesse}}{{Anders} \&
  {Grevesse}}{1989}]{Anders1989}
{Anders} E.,  {Grevesse} N.,  1989, \gca, 53, 197

\bibitem[\protect\citeauthoryear{{Churazov}, {Gilfanov}, {Forman} \&
  {Jones}}{{Churazov} et~al.}{1999}]{Churazov1999}
{Churazov} E.,  {Gilfanov} M.,  {Forman} W.,    {Jones} C.,  1999, \apj, 520,
  105

\bibitem[\protect\citeauthoryear{{De Luca} \& {Molendi}}{{De Luca} \&
  {Molendi}}{2004}]{DeLuca2004}
{De Luca} A.,  {Molendi} S.,  2004, \aap, 419, 837

\bibitem[\protect\citeauthoryear{{Diehl} \& {Statler}}{{Diehl} \&
  {Statler}}{2006}]{Diehl2006}
{Diehl} S.,  {Statler} T.~S.,  2006, \mnras, 368, 497

\bibitem[\protect\citeauthoryear{{Dupke} \& {Bregman}}{{Dupke} \&
  {Bregman}}{2001}]{Dupke2001}
{Dupke} R.~A.,  {Bregman} J.~N.,  2001, \apj, 562, 266

\bibitem[\protect\citeauthoryear{{Fabian}, {Sanders}, {Taylor} \&
  {Allen}}{{Fabian} et~al.}{2005}]{Fabian2005}
{Fabian} A.~C.,  {Sanders} J.~S.,  {Taylor} G.~B.,    {Allen} S.~W.,  2005,
  \mnras, 360, L20

\bibitem[\protect\citeauthoryear{{Forman}, {Jones}, {Churazov}, {Markevitch},
  {Nulsen}, {Vikhlinin}, {Begelman}, {B{\"o}hringer}, {Eilek}, {Heinz},
  {Kraft}, {Owen} \& {Pahre}}{{Forman} et~al.}{2007}]{Forman2007}
{Forman} W.,  {Jones} C.,  {Churazov} E.,  {Markevitch} M.,  {Nulsen} P.,
  {Vikhlinin} A.,  {Begelman} M.,  {B{\"o}hringer} H.,  {Eilek} J.,  {Heinz}
  S.,  {Kraft} R.,  {Owen} F.,    {Pahre} M.,  2007, \apj, 665, 1057

\bibitem[\protect\citeauthoryear{{Furusho}, {Yamasaki}, {Ohashi}, {Shibata},
  {Kagei}, {Ishisaki}, {Kikuchi}, {Ezawa} \& {Ikebe}}{{Furusho}
  et~al.}{2001}]{Furusho2001}
{Furusho} T.,  {Yamasaki} N.~Y.,  {Ohashi} T.,  {Shibata} R.,  {Kagei} T.,
  {Ishisaki} Y.,  {Kikuchi} K.,  {Ezawa} H.,    {Ikebe} Y.,  2001, \pasj, 53,
  421

\bibitem[\protect\citeauthoryear{{Hasinger}, {Miyaji} \& {Schmidt}}{{Hasinger}
  et~al.}{2005}]{Hasinger2005}
{Hasinger} G.,  {Miyaji} T.,    {Schmidt} M.,  2005, \aap, 441, 417

\bibitem[\protect\citeauthoryear{{Henry}, {Finoguenov} \& {Briel}}{{Henry}
  et~al.}{2004}]{Henry2004}
{Henry} J.~P.,  {Finoguenov} A.,    {Briel} U.~G.,  2004, \apj, 615, 181

\bibitem[\protect\citeauthoryear{{Kalberla}, {Burton}, {Hartmann}, {Arnal},
  {Bajaja}, {Morras} \& {P{\"o}ppel}}{{Kalberla} et~al.}{2005}]{LABsurvey}
{Kalberla} P.~M.~W.,  {Burton} W.~B.,  {Hartmann} D.,  {Arnal} E.~M.,  {Bajaja}
  E.,  {Morras} R.,    {P{\"o}ppel} W.~G.~L.,  2005, \aap, 440, 775

\bibitem[\protect\citeauthoryear{{Leccardi} \& {Molendi}}{{Leccardi} \&
  {Molendi}}{2008}]{Leccardi2008}
{Leccardi} A.,  {Molendi} S.,  2008, \aap, 486, 359

\bibitem[\protect\citeauthoryear{{Lucey}, {Currie} \& {Dickens}}{{Lucey}
  et~al.}{1986}]{Lucey1986}
{Lucey} J.~R.,  {Currie} M.~J.,    {Dickens} R.~J.,  1986, \mnras, 222, 427

\bibitem[\protect\citeauthoryear{{Mazzotta}, {Rasia}, {Moscardini} \&
  {Tormen}}{{Mazzotta} et~al.}{2004}]{Mazzotta2004}
{Mazzotta} P.,  {Rasia} E.,  {Moscardini} L.,    {Tormen} G.,  2004, \mnras,
  354, 10

\bibitem[\protect\citeauthoryear{{Nevalainen}, {David} \&
  {Guainazzi}}{{Nevalainen} et~al.}{2010}]{Nevalainen2010}
{Nevalainen} J.,  {David} L.,    {Guainazzi} M.,  2010, \aap, 523, A22

\bibitem[\protect\citeauthoryear{{Ota}, {Fukazawa}, {Fabian}, {Kanemaru},
  {Kawaharada}, {Kawano}, {Kelley}, {Kitaguchi}, {Makishima}, {Matsushita},
  {Murase}, {Nakazawa}, {Ohashi}, {Sanders}, {Tamura} \& {Urata}}{{Ota}
  et~al.}{2007}]{Ota2007}
{Ota} N.,  {Fukazawa} Y.,  {Fabian} A.~C.,  {Kanemaru} T.,  {Kawaharada} M.,
  {Kawano} N.,  {Kelley} R.~L.,  {Kitaguchi} T.,  {Makishima} K.,  {Matsushita}
  K.,  {Murase} K.,  {Nakazawa} K.,  {Ohashi} T.,  {Sanders} J.~S.,  {Tamura}
  T.,    {Urata} Y.,  2007, \pasj, 59, 351

\bibitem[\protect\citeauthoryear{{Sanders} \& {Fabian}}{{Sanders} \&
  {Fabian}}{2006}]{Sanders2006}
{Sanders} J.~S.,  {Fabian} A.~C.,  2006, \mnras, 371, 1483

\bibitem[\protect\citeauthoryear{{Simionescu}, {B{\"o}hringer}, {Br{\"u}ggen}
  \& {Finoguenov}}{{Simionescu} et~al.}{2007}]{Simionescu2007}
{Simionescu} A.,  {B{\"o}hringer} H.,  {Br{\"u}ggen} M.,    {Finoguenov} A.,
  2007, \aap, 465, 749

\bibitem[\protect\citeauthoryear{{Smith}, {Brickhouse}, {Liedahl} \&
  {Raymond}}{{Smith} et~al.}{2001}]{Smith2001}
{Smith} R.~K.,  {Brickhouse} N.~S.,  {Liedahl} D.~A.,    {Raymond} J.~C.,
  2001, \apjl, 556, L91

\bibitem[\protect\citeauthoryear{{Snowden}, {Mushotzky}, {Kuntz} \&
  {Davis}}{{Snowden} et~al.}{2008}]{Snowden2008}
{Snowden} S.~L.,  {Mushotzky} R.~F.,  {Kuntz} K.~D.,    {Davis} D.~S.,  2008,
  \aap, 478, 615

\bibitem[\protect\citeauthoryear{{Urban}, {Werner}, {Simionescu}, {Allen} \&
  {B{\"o}hringer}}{{Urban} et~al.}{2011}]{Urban2011}
{Urban} O.,  {Werner} N.,  {Simionescu} A.,  {Allen} S.~W.,    {B{\"o}hringer}
  H.,  2011, \mnras, 414, 2101

\bibitem[\protect\citeauthoryear{{Walker}, {Fabian}, {Sanders}, {Simionescu} \&
  {Tawara}}{{Walker} et~al.}{2013}]{Walker2013_Centaurus}
{Walker} S.~A.,  {Fabian} A.~C.,  {Sanders} J.~S.,  {Simionescu} A.,
  {Tawara} Y.,  2013, \mnras, 432, 554

\end{thebibliography}
